\documentclass[journal]{IEEEtran}

\ifCLASSINFOpdf
  \usepackage[pdftex]{graphicx}
\else
\fi

\usepackage{amsmath}
\usepackage{amssymb}

\begin{document}

\title{MRDust: Wireless Implant Data Uplink \& Localization via Magnetic Resonance Image Modulation}

\author{Biqi Rebekah Zhao,~\IEEEmembership{Graduate Student Member,~IEEE},
        Alexander Chou,~\IEEEmembership{Graduate Student Member, IEEE},
        Robert Peltekov,
        Elad Alon,~\IEEEmembership{Fellow, IEEE},
        Chunlei Liu,
        Rikky Muller*,~\IEEEmembership{Senior Member, IEEE},
        and~Michael Lustig*% <-this % stops a space
\vspace{-6mm}
\thanks{Biqi Rebekah Zhao, Alexander Chou, Robert Peltekov, Elad Alon, and Michael Lustig are with the Department
of Electrical Engineering and Computer Sciences, University of California, Berkeley,
CA, 94720, USA e-mail: (rebekah$\_$zhao@berkeley.edu).}
\thanks{Chunlei Liu is with the Department
of Electrical Engineering and Computer Sciences, University of California, Berkeley,
CA, 94720, and also with the Helen Wills Neuroscience Institute, Berkeley, CA, 94720}
\thanks{Rikky Muller is with the Department
of Electrical Engineering and Computer Sciences, University of California, Berkeley,
CA, 94720, and also with the Weil Neurohub, Berkeley, CA, 94720}
\thanks{*These authors contributed equally as last authors.}
\thanks{This work has been submitted to the IEEE for possible publication. Copyright may be transferred without notice, after which this version may no longer be accessible.}
}
%\markboth{Transactions on Biomedical Circuits and Systems}%

\maketitle 
\vspace{-2em}

\begin{abstract}
Magnetic resonance imaging (MRI) exhibits rich and clinically useful endogenous contrast mechanisms, which can differentiate soft tissues and are sensitive to flow, diffusion, magnetic susceptibility, blood oxygenation level, and more. However, MRI sensitivity is ultimately constrained by Nuclear Magnetic Resonance (NMR) physics, and its spatiotemporal resolution is limited by SNR and spatial encoding. 
On the other hand, miniaturized implantable sensors offer highly localized physiological information, yet communication and localization can be challenging when multiple implants are present. This paper introduces the MRDust, an active ``contrast agent" that integrates active sensor implants with MRI, enabling the direct encoding of highly localized physiological data into MR images to augment the anatomical images. MRDust employs a micrometer-scale on-chip coil to actively modulate the local magnetic field, enabling MR signal amplitude and phase modulation for digital data transmission. Since MRI inherently captures the anatomical tissue structure, this method has the potential to enable simultaneous data communication, localization, and image registration with multiple implants. This paper presents the underlying physical principles, design tradeoffs, and design methodology for this approach. To validate the concept, a 900 $\times$ 990 $\mu$m$^2$ chip was designed using TSMC 28 nm technology, with an on-chip coil measuring 630 $\mu$m in diameter. The chip was tested with custom hardware in an MR750W GE3T MRI scanner. Successful voxel amplitude modulation is demonstrated with Spin-Echo Echo-Planar-Imaging (SE-EPI) sequence, achieving a contrast-to-noise ratio (CNR) of 25.58 with a power consumption of 130 $\mu$W. 

\end{abstract}


\IEEEpeerreviewmaketitle

\section{Introduction}

\begin{figure}[t!]
\centerline{\includegraphics[width=8.2cm]{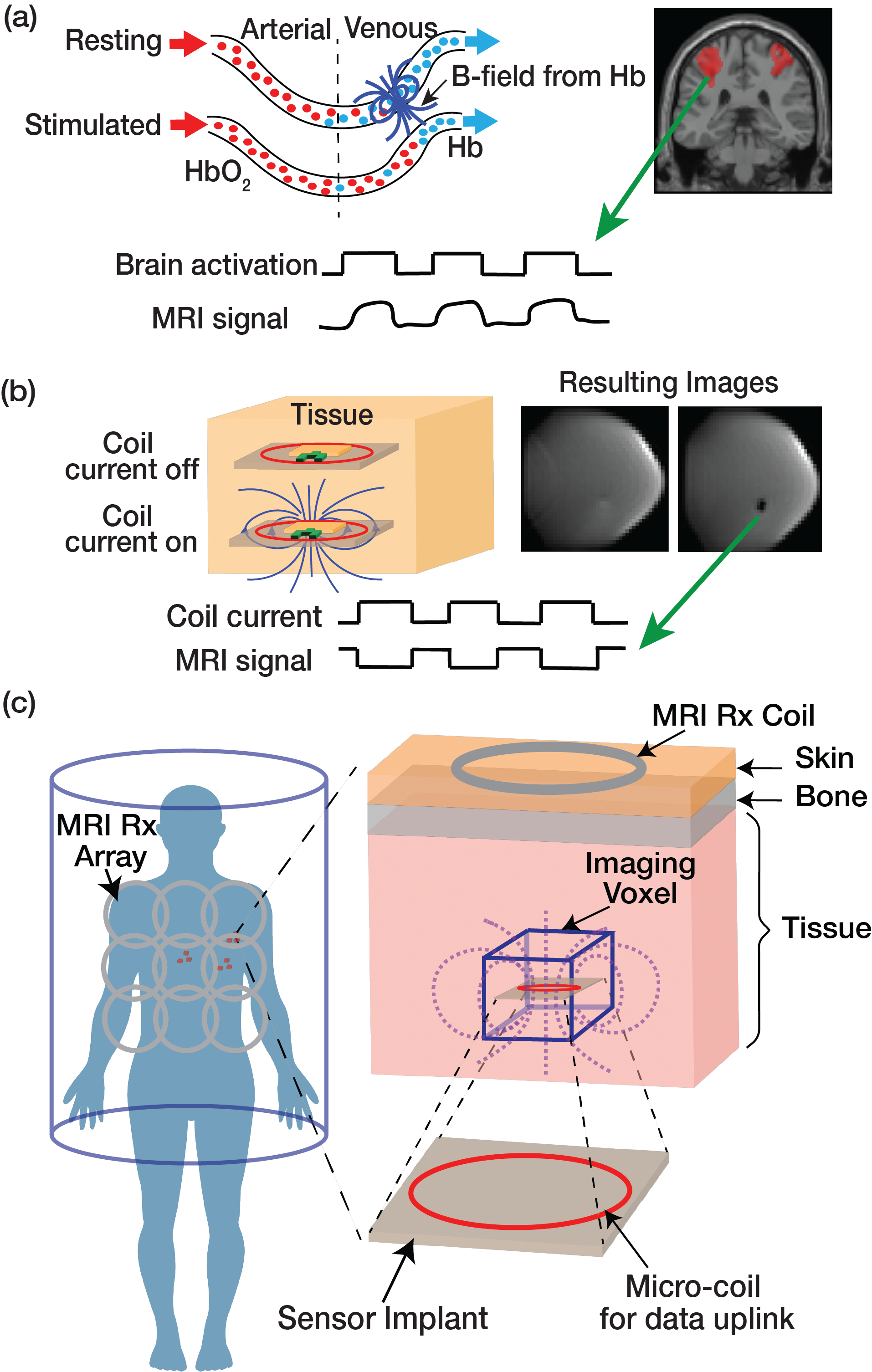}}
\caption{(a) Brain activation drives changes in magnetic susceptibility which causes local modulation of the magnetic field, giving rise to signal modulation in BOLD fMRI. (b) Current flowing through the on-chip micro-coil generates local magnetic field changes that result in voxel amplitude modulation in the images. (c) Concept of MRDust: An on-chip coil is integrated with a sensor implant to act as a ``smart contrast agent". The coil modulates the local magnetic field to encode sensor data in MR images so that the images are augmented with physiological information.\vspace{-2em}}
\label{f1}
\end{figure}

\IEEEPARstart{M}{agnetic} resonance imaging (MRI) is a versatile medical imaging modality that not only captures detailed anatomical structures of soft tissues, but also monitors physiological processes such as blood flow, and biomarkers for brain activity and tissue metabolism. The power of MRI lies in the generation of contrast. In static imaging, tissue-specific parameters, including proton density, $T_1$, and $T_2$ relaxation times, are combined with imaging techniques and contrast agents to produce distinct contrasts that differentiate various tissue types \cite{nitz1999contrast}. Beyond static imaging, the sensitivity of the MR signal to subtle changes in the magnetic field enables the detection of dynamic physiological phenomena. Functional parameters such as diffusion, perfusion, and flow influence contrast, providing insights into physiological processes. For example, in blood oxygenation level-dependent functional MRI (BOLD fMRI), increased brain activity in response to external stimuli such as visual and audio cues leads to changes in blood oxygenation, volume, and flow.
These hemodynamic changes induce local magnetic field fluctuations that, in turn, modulate the MR signal \cite{glover2011overview} (Fig. 1(a)). As a result, regions of the brain engaged during specific tasks can be identified directly from the MR images. However, a limitation of this approach is that the detected signal changes reflect secondary hemodynamic responses to neural activity, rather than the neural activity itself.

Over the past decade, advancements in nanofabrication and integrated circuits (IC) technologies have driven significant progress in implantable electronics, enabling the development of miniaturized wireless sensors for continuous monitoring of biophysical and bioelectrical parameters \cite{yogev2023current}. These sensors offer localized measurements with relatively high temporal resolution while maintaining a minimally invasive profile. For instance, \cite{ghanbari2019sub} presents a 0.25 $\text{mm}^2$ device for direct action potential recording with up to 5 kHz bandwidth. Other implementations include implants for glucose monitoring \cite{mujeeb2019novel}, temperature sensing \cite{shi2021application}, and pressure sensing \cite{weber2018miniaturized}. These implantable sensors report similar sensitivity and temporal resolution to their state-of-the-art wired counterparts while achieving much smaller form factors and power consumption. However, challenges arise when multiple implants are deployed at different locations, particularly in managing communication through a single channel and accurately localizing the implants. To overcome communication hurdles, sophisticated protocols for different power and communication modalities such as those proposed in \cite{lee2021neural}, \cite{alamouti2020high} and \cite{lee202535} have been developed. Similarly, accurate localization of implants necessitates custom hardware solutions, as demonstrated in \cite{monge2017localization}.

Combining MRI with wireless sensor implants could create a synergistic approach that leverages the strength of both modalities, providing detailed anatomical imaging alongside localized physiological data. Distributed sensor implants, used in conjunction with MRI, could offer physiological measurements embedded within MR images, enabling simultaneous data acquisition across a large spatial area and direct visualization of implant locations (Fig 1. (c)). The concept of using MRI contrast to communicate biophysical data was previously explored in \cite{hai2019wireless}, where millimeter-sized inductively coupled resonators were tuned or detuned to encode data into MR images. However, this method has limitations: 1) the resonators are difficult to further miniaturize due to the resonant frequency constraints of MR scanners, and 2) the passive nature of the data uplink restricts the flexibility of data encoding and the types of sensors that can be used. Therefore, an active data encoding method with a smaller form factor could be beneficial.

This article proposes the ``MRDust" --- a ``smart contrast agent" that enables active encoding of data in MR images to augment anatomical information with physiological insights. In this method, a micrometer-scaled coil on an IC chip modulates the current level in a binary fashion,  causing local magnetic field changes that appear as MR signal amplitude and phase modulation (Fig. \ref{f1} (b)). Such modulation can be used to encode digital bits representing measured physiological data from a miniature sensor implant. The on-chip coil, if integrated with wireless energy harvesting, sensors, and sensor acquisition circuits, has the potential to monitor and transmit local physiological information in deep tissues. Since the signal from each implant appears simultaneously on MR images, if multiple implants are present, parallel data readout, implant localization, and image registration could be possible. A network of MRDust implants could offer benefits beyond existing diagnostic solutions. For example, deploying a diverse set of sensors - capable of detecting cancer biomarkers such as oxygenation, pH and specific chemicals - near a tumor resection site after surgery could complement MRI scans by providing detailed insights into the tumor microenvironment and pinpointing locations of abnormality, enhancing the monitoring of tumor recurrence and progression \cite{chen2023hypoxic}\cite{hosonuma2023association}.

In this paper, we focus only on the MR-based data uplink method, as various methods on wireless energy harvesting, downlink communication protocol, and miniaturized implantable sensor design have been explored in prior work \cite{singer2021wireless} \cite{karimi2021wireless} \cite{yogev2023current}. The paper further explains and expands the work presented in \cite{zhao2024mrdust} and is organized as follows: Section II explains the physical principles of MRI. Section III provides a detailed explanation of how to create MR imaging contrast synchronized with MRI pulse sequences using a micro-coil, and its design trade-offs and design procedure. Section IV outlines the design of a proof-of-concept prototype and the test setup, with a focus on tackling the challenges of testing electronics in the MRI environment. Section V presents \emph{in vitro} experimental results from the prototype, followed by conclusions and discussions in section VI.

\section{Magnetic Resonance Imaging Basics}

The necessary background in MR physics \cite{haacke1999magnetic}, \cite{buxton2009introduction} is presented in this section, providing the foundation for MRDust and serving to introduce its underlying concepts.

\subsection{Magnetic Resonance Physical Principles}
\subsubsection{Magnetization}
The MR signal in the body comes from the spins of hydrogen nuclei, mostly in water. The presence of a strong magnetic field $\vec{B}=B_0\hat{z}$ gives rise to a net magnetization vector $\vec{M}=M_0\hat{z}$ in a volume

\begin{equation}
M_0 = \frac{N \gamma ^2\hbar ^2 I_z(I_z+1)B_0 V}{3kT},
\label{eq1}
\end{equation}
\noindent where $N$ is the number of nuclear spins per unit volume, $V$ is the volume, $\hbar$ is the reduced Planck constant, and $I_z = \frac{1}{2}$ is the \textit{z}-component of the nuclear spin angular momentum \cite{abragam1961principles}. 

The magnetization exhibits resonance at the Larmor frequency $\omega_0$ given by 
\begin{equation}
\omega_0 = \gamma B_0,
\label{eq2}
\end{equation}
where $\gamma$ is the gyromagnetic ratio ($\gamma/2\pi \approx 42.58~\text{MHz}/\text{T}$ for hydrogen).

\begin{figure}[t!]
\centerline{\includegraphics[width=8.2cm]{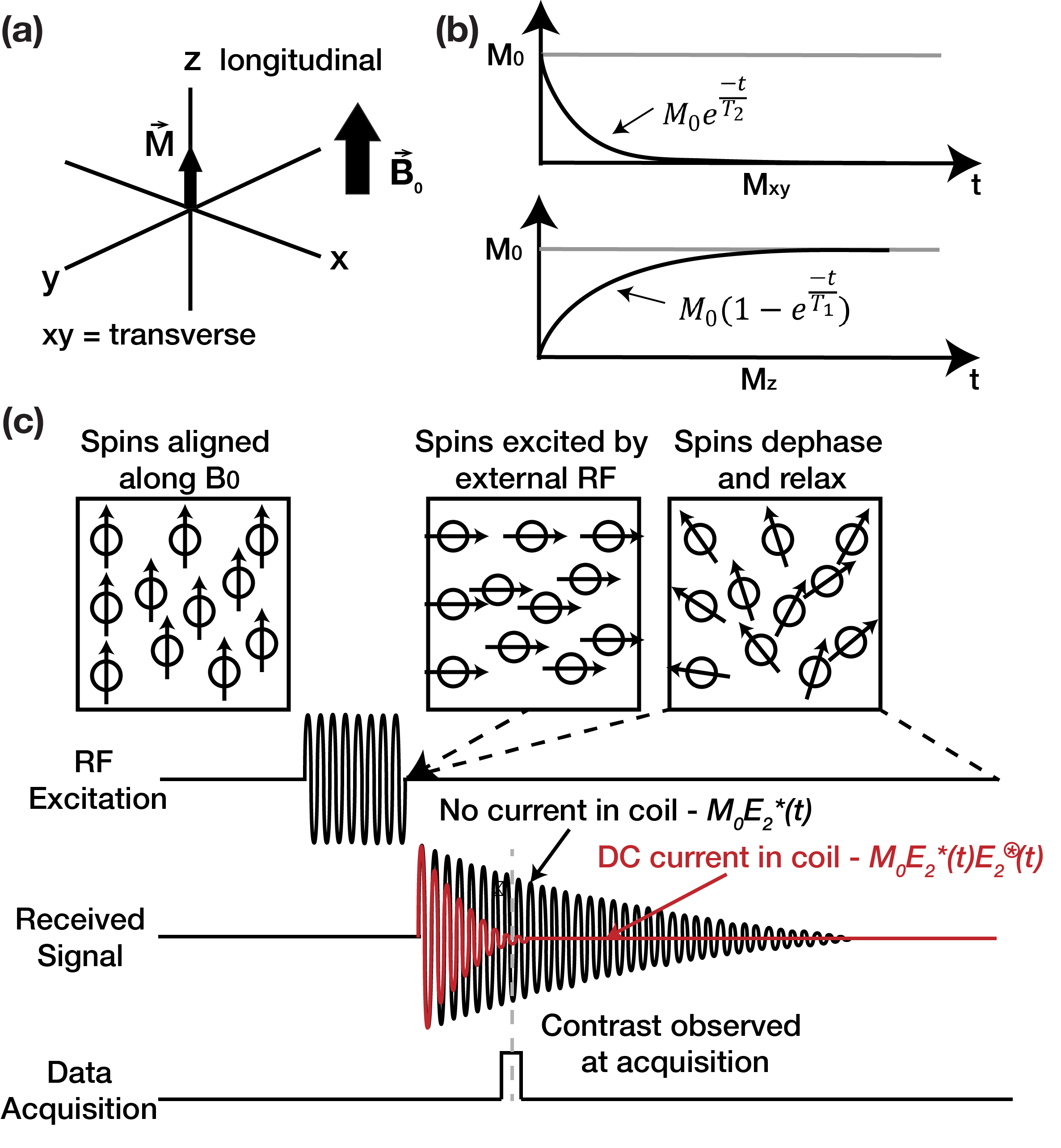}}
\caption{(a) In a strong magnetic field $\vec{B_0}$, spins are aligned in the longitudinal direction. (b) After an RF excitation, the longitudinal and transverse magnetization go through $T_1$ recovery and $T_2$ relaxation, respectively. (3) To actively encode data in an image, a current-carrying micro-coil induces magnetic field inhomogeneities that cause spins to dephase more rapidly, reducing the effective decay time constant and, in turn, the received signal amplitude. Amplitude contrast is observed at acquisition between current-on and current-off states.\vspace{-2em}}
\label{f3}
\end{figure}
\vspace{-2mm}
\subsubsection{Excitation, Precession and Reception}

The magnetization can be excited by applying a radio-frequency (RF) field $\vec{B_1}$ at the Larmor frequency in the transverse ($\hat{x}-\hat{y}$) plane. The magnetization tips away from the longitudinal ($\hat{z}$) axis towards the transverse plane, creating a transverse magnetization component, which precesses around $\hat{z}$ at $\omega_0$ and is conveniently described by the phasor $M_{xy}$. The level of excitation is measured by the tip-angle, where a $90^\circ$ RF pulse will result in the entirety of the magnetization in the $M_{xy}$ component.

The $M_{xy}$ component can be detected through induction by placing a receiver RF-coil tuned to $\omega_0$ near the sample. The change in flux induces an electromotive force (EMF) in the receiver coil, which through Faraday’s law of induction and the principle of reciprocity \cite{hoult1978nmr} can be presented as 

\begin{equation}
\delta \epsilon = -\frac{\partial}{\partial t}\{\vec{B_p}\cdot \vec{M}\} \delta V_s,
\label{eq3}
\end{equation}
where $\vec{B_p}$ is the magnetic field produced by the receiver coil at the sample location per unit current and $V_s$ is an elementary sample volume.

It can be shown that the induced signal, or the voltage on the MR receiver coil $S_r(t)\propto M_{xy}e^{j\omega_0 t}$ is proportional to the transverse magnetization. For simplicity, we demodulate the signal by $\omega_0$ and consider the baseband signal $S_{bb}(t)=M_{xy}(t)$ for the rest of the analysis.

\subsubsection{Relaxation}

 Once excited, the magnetization also exhibits relaxation. The transverse component $M_{xy}$ decays, while the longitudinal component $M_z$ recovers to equilibrium. These processes occur with characteristic exponential relaxation time constants $T_2$ and $T_1$ respectively as shown in Fig. \ref{f3} (b). Following a $90^\circ$ RF excitation, the baseband magnetization can be expressed as:

\begin{equation}
M_{z}(t) = M_0(1-e^{-\frac{t}{T_1}}),
\label{eq4}
\end{equation}

\begin{equation}
M_{xy}(t) = M_0e^{-\frac{t}{T_2}},
\label{eq5}
\end{equation}
which results in 
\begin{equation}
S_{bb}(t) = M_0e^{-\frac{t}{T_2}}.
\label{eq6}
\end{equation}

\begin{figure*} [t!]
\includegraphics[width=17cm]{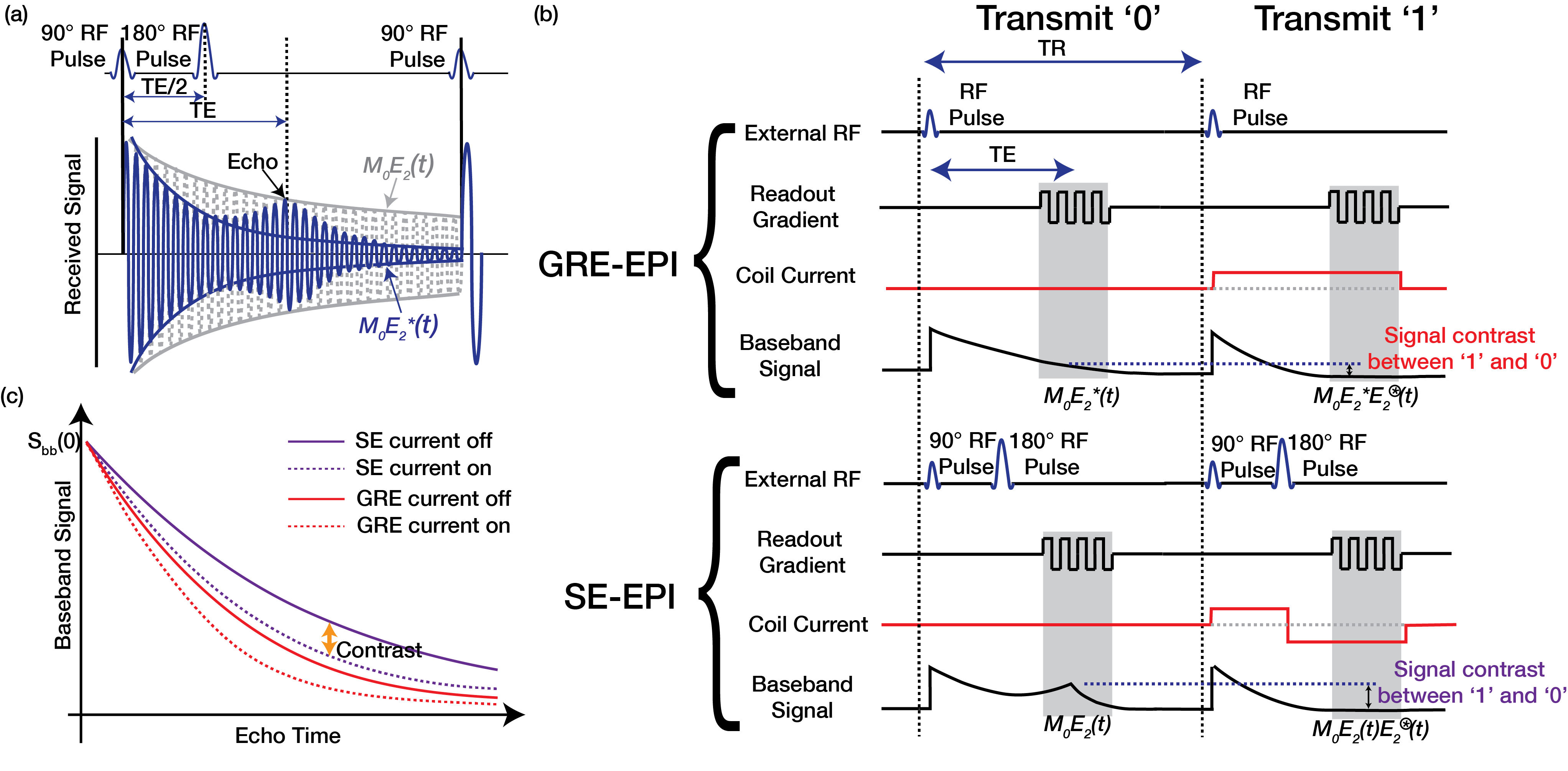}
\centering
\caption{(a) Illustration of an SE pulse sequence.
(b) Synchronization of the micro-coil current to GRE-EPI and SE-EPI sequences to encode digital ``0" and ``1".  (c) Received baseband signal contrast for GRE and SE between current-on and current-off states as echo time increases.\vspace{-2em}}
\label{f4}
\end{figure*}

\subsubsection{Intra-voxel Field Inhomogeneity}

Due to the heterogeneity of structures and magnetic susceptibility in tissue, natural magnetic field inhomogeneities, denoted as $\Delta B_{\mathrm{inhom}}$, are present. The inhomogeneities within a voxel cause the magnetization to lose coherence and dephase, resulting in additional signal decay. This decay is typically approximated as exponential and the observed signal for a $90^\circ$ excitation is therefore 
\begin{equation}
	S_{bb}(t) = M_0e^{-\frac{t}{T_2^*}} = M_0e^{-(\frac{t}{T_2}+\frac{t}{T_2'})},
\label{eq7}
\end{equation}
where $\frac{1}{T_2’}\propto \Delta B_\mathrm{inhom}$, and ${T_2^*}$ is the actual effective decay time constant, and is a commonly used concept in MRI \cite{chavhan2009principles}. For simplicity, in the rest of the paper, we define $E_2(t) = e^{-\frac{t}{T_2}}$, $E_2^*(t) = e^{-\frac{t}{T_2^*}}$, and so on. $S_{bb}(t)$ including inhomogeneity can then be represented as 
\begin{equation}
S_{bb}(t) = M_0E_2^*(t).
\label{eq8}
\end{equation}

Inhomogeneity can also be induced artificially by an electric current carrying micro-coil. By Bio-Savart Law, a magnetic field $\Delta B_\mathrm{ind}$ is generated around a coil when current flows through it, creating a similar effect to the natural inhomogeneity, which leads to additional exponential relaxation. We define the decay time constant associated with $\Delta B_\mathrm{ind}$ as $T_2^{\circledast}$, where $\frac{1}{T_2^{\circledast}}\propto\Delta B_\mathrm{ind}$. The baseband signal with current flowing through the micro-coil becomes 

\begin{equation}
	S_{bb,on}(t) = M_0E_2^*(t)E_2^{\circledast}(t).
\label{eq9}
\end{equation}

To enable the creation of active contrast in a voxel, the effective decay time constant can be altered by turning the current in the micro-coil on and off. When no current flows through the micro-coil, the $E_2^{\circledast}(t) = 1$, and $S_{bb,off}$ is the same as in (\ref{eq8}). Since the current-on and current-off states exhibit different signal decay time constant, at the time of data acquisition, the two states produce different amplitudes in the received signal. This contrast in signal amplitude can be used to achieve binary data encoding (Fig. \ref{f3} (c)).

\vspace{-2mm}
\subsection{Pulse Sequences}

An MRI pulse sequence is a pre-programmed set of RF pulses, gradient pulses, and timing parameters that control how an MRI system acquires data. Example sequences are shown in Fig. \ref{f4} (b). A sequence begins with an RF pulse to excite the spins. After a delay, signal contrasts evolve due to tissue-specific relaxation properties. Magnetic field gradients spatially encode the signals, which is captured by the receiver chain. The time at which data acquisition happens is called echo time (TE), and the interval between successive excitations is called the repetition time (TR). We describe the two types of pulse sequences used for this application: Gradient Echo (GRE) and Spin Echo (SE).
\subsubsection{Gradient Echo (GRE)}
In GRE, a single RF pulse smaller than 90$^\circ$ is typically used to excite the spins \cite{markl2012gradient}. After excitation, the spins dephase and relax with a time constant  $T_2^*$. As a result, the received baseband signal at data acquisition is $S_{bb}(TE) = M_0E_2^*(TE)$. Therefore, GRE is sensitive to the natural field inhomogeneities, and signal loss is severe for long TE. 
\subsubsection{Spin Echo (SE)}
Unlike GRE, the SE sequence is insensitive to the natural field inhomogeneities \cite{jung2013spin}. A spin echo is generated using a pair of RF pulses. At the beginning of each TR, a 90$^\circ$ pulse excites the spins. The spins initially dephase and relax with a time constant $T_2^*$, similar to GRE. At TE/2, a 180$^\circ$ RF pulse is applied, negating the spin phases. The spins then rephase and form an ``echo" at TE on the $T_2$ relaxation curve  (Fig. \ref{f4} (a)). After the echo, the spins dephase again and resume following the $T_2^*$ relaxation curve. Therefore, at the time of data acquisition TE, the received baseband signal becomes $S_{bb}(TE) = M_0E_2(TE)$ and is unaffected by the natural field inhomogeneities.

\subsubsection{Echo Planar Imaging (EPI)} Similar to fMRI, high temporal resolution is desirable for the proposed data encoding approach. Therefore, we use a fast sequence called Echo-Planar Imaging (EPI) to acquire an image per TR \cite{norris2015pulse}. EPI can be combined with GRE and SE to form GRE-EPI and SE-EPI sequences.

\section{Data Encoding in Magnetic Resonance Images}

\subsection{Active Signal Modulation in MRI}
To efficiently encode data into images, we synchronize the operation of the micro-coil with EPI sequences. The approaches to synchronize with GRE-EPI and SE-EPI are different. Fig. \ref{f4} (b) top shows an implementation synchronized with GRE-EPI. Binary encoding is achieved by modulating the micro-coil current during each TR. To encode a digital ``1", the current is turned on immediately after the RF excitation and remains on until the end of data acquisition. The received voxel amplitude $S_{bb,on}(TE) = M_0E_2^*(TE)E_2^{\circledast}(TE)$. When the micro-coil current is off, voxel intensity remains unaffected, representing a digital ``0", and  $S_{bb,off}(TE)=M_0E_2^*(TE)$ is received. 
\begin{figure*}[t!]
\centering
\includegraphics[width=15cm]{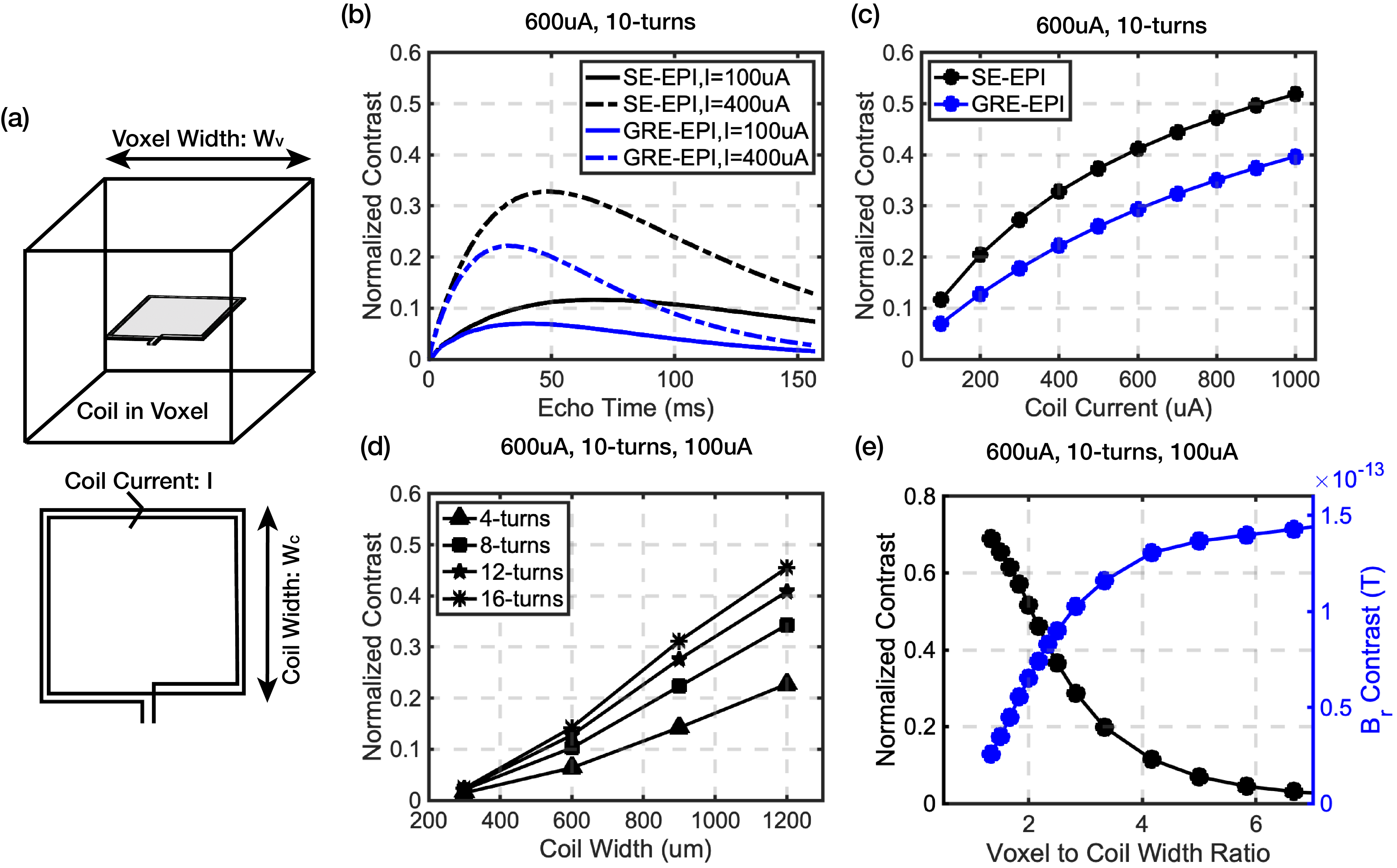}
\caption{(a) Illustration of coil placement in an MR imaging voxel and optimization parameters in simulation. (b) Normalized contrast for GRE-EPI and SE-EPI over echo time. (c) Maximum normalized contrast for GRE-EPI and SE-EPI at different current levels. (d) Maximum normalized contrast for SE-EPI with different coil widths and number of turns. (e) Changes in normalized contrast and detectable $\vec{B_r}$ field contrast with increasing voxel size to coil size ratio. \vspace{-2em}}
\label{f5}
\end{figure*}

However, the induced inhomogeneity competes with the natural inhomogeneity, limiting the detectable contrast. To mitigate this issue, we could leverage SE-EPI, but the 180$^\circ$ refocusing pulse would refocus the coil-induced inhomogeneities, which is undesirable. Instead, the micro-coil current is reversed during the refocusing pulse, flipping the $\Delta B_{\text{ind}}$ direction. As a result, the effects of the 180$^\circ$ refocusing pulse and the $\Delta B_{\text{ind}}$ reversal cancel each other for spins around the micro-coil, and they continue to dephase while other sources of dephasing refocus. As shown in Fig \ref{f4}. (b) bottom, encoding a digital ``1" involves turning on the micro-coil current after the 90$^\circ$ excitation, flipping its direction during the refocusing pulse, and turning it off after image acquisition. For the SE-EPI pulse sequence, $S_{bb,on}(TE) = M_0E_2(TE)E_2^{\circledast}(TE)$ and $S_{bb,off}(TE) = M_0E_2(TE)$. This method generates a higher contrast between current-on and current-off states compared to the GRE-EPI sequence.
\vspace{-2mm}
\subsection{Design Considerations}
In the proposed approach, several parameters affect the modulation efficiency of the micro-coil. The key to creating a larger contrast is to reduce $E_2^{\circledast}(t)$, which indicates maximizing $\Delta B_{\text{ind}}$. A larger $\Delta B_{\text{ind}}$ can be achieved by increasing the micro-coil current, number of turns, or coil size. However, these would lead to greater implant size and higher power consumption. Modulation efficiency is also affected by pulse sequence parameters, including voxel size, TE, and TR. To better understand these dependencies, simulations of the effects of multiple parameters on modulation efficiency are shown in Figures \ref{f5}, \ref{f6}, and \ref{f7}.

Temporal contrast-to-noise ratio (tCNR) is a common metric in fMRI to evaluate the quality of fMRI methods in identifying contrast due to brain activity \cite{geissler2007contrast}. For implantable applications, the aim is to find the minimum tCNR that results in sufficient contrast while minimizing the micro-coil current and area. In order to achieve this goal, the effect of each parameter on tCNR was simulated. Since MR image noise depends on various factors, including acquisition hardware and the imaging sequence, we instead assume a known noise level and focus on maximizing the contrast between digital ``1" and ``0" states. The contrast is normalized to the baseband signal immediately after excitation ($S_{bb}(0)$) to eliminate the dependency on MR pulse sequence parameters. This parameter is defined as normalized contrast $C_n$. Fig. \ref{f4} (c) illustrates the detected baseband signal as TE increases for GRE and SE sequences. The contrast level is proportional to $S_{bb}(0)$. $C_n$ is thereby defined as

\begin{equation}
C_n = \frac{S_{bb,off}(t) - S_{bb,on}(t)}{S_{bb}(0)}.
\label{eq12}
\end{equation}

The simulation setup is shown in Fig. \ref{f5} (a). A square-spiral-shaped current-carrying micro-coil with $N_t$ turns, an outer width of $W_c$, and a current $I$ is positioned in the center of a square voxel of width $W_v$. All simulations were performed using MATLAB. The magnetic field generated by the micro-coil was simulated with the BSmag Toolbox \cite{BSmag}. A single voxel was quantized into a million equally-spaced spins, and a Bloch simulator from Stanford University was then used to model the received MR signal. Table \ref{tab:t1} lists the $T_1$, $T_2$, and $T_2^*$ values used in the simulation, which are based on literature values for white matter at 3T \cite{wansapura1999nmr}. The voxel size is defined to be 2$\times$2$\times$2 mm$^3$, which is typical in fMRI. The following parameters that affect contrast are considered:

\begin{table}
\centering
 \caption{MR parameters for Bloch simulation at 3T}
 \label{tab:t1}
\begin{tabular} { |c||c|c|c|  }

 \hline
 Tissue& T1 (ms) &T2 (ms) &T2* (ms)\\
 \hline
 White Matter & 832   &80&   44.7\\

 \hline

 \end{tabular}
 \vspace{-2em}
\end{table}

\subsubsection{Echo Time (TE)}
Fig. \ref{f5} (b) shows the relationship between $C_n$ and TE for current levels 100 $\mu$A and 400 $\mu$A in a 600 $\mu$m 10-turn micro-coil with 10 $\mu$m spacing. For both GRE and SE, a higher current level results in higher contrast, with SE showing an overall higher contrast than GRE. The contrast increases over time until the overall magnetization decreases to the point that the contrast starts decreasing. The optimal TE values maximizing $C_n$ are approximately 40 ms for GRE and 65 ms for this particular design. 

\subsubsection{Coil Current}
Fig. \ref{f5} (c) illustrates how contrast varies with coil current. For the same micro-coil setup as Fig. \ref{f5} (b), the largest achievable $C_n$ is plotted for different current levels. The contrast increases with increasing current, but the relationship is nonlinear. This indicates that multi-level amplitude modulation could be explored, but the nonlinearity needs to be compensated for.

\begin{figure*}[t!]
\includegraphics[width=\textwidth]{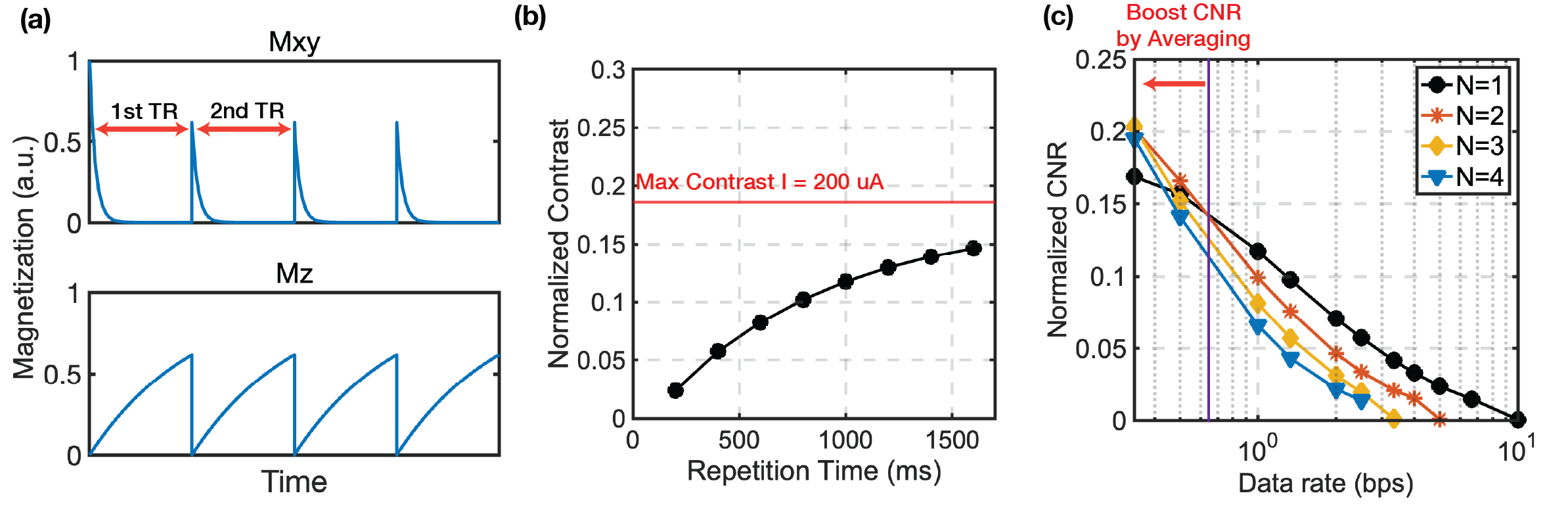}
\caption{(a) Illustration of the longitudinal and transverse magnetization in a few consecutive TRs. (b) Maximum normalized contrast with the same coil design in Fig. \ref{f5} for different repetition times. The red line represents the maximum achievable contrast at the same TE when TR = $\infty$. (c) The effect of averaging and simply increasing TR on normalized CNR for different data rates.\vspace{-2em}}
\label{f6}
\end{figure*}

\subsubsection{Coil Design}
Fig. \ref{f5} (d) examines $C_n$ scaling with coil size and the number of coil turns in the SE pulse sequence. The increase in contrast is more significant by increasing coil size rather than the number of turns. This is because the coil design is limited to 2-D, meaning each additional turn has to spiral inward, creating a loop with a smaller width. In process technologies with several thick metal layers, it is possible to design the micro-coil on multiple layers to enhance contrast without increasing area.

\subsubsection{Voxel Size}
For a chosen micro-coil design, the relationship between coil size and voxel size also affects the modulation efficiency. The black curve in Fig. \ref{f5} (e) illustrates how $C_n$ varies with the voxel to coil width ratio for a coil with 10 turns, 600 $\mu$m width, and 100 $\mu$A current. As the voxel size becomes significantly larger than the on-chip coil, the effect of the micro-coil becomes spatially confined, and $C_n$ decreases substantially. 

However, $C_n$ alone does not fully capture the effects of voxel size, since its normalizing factor $S_{bb}(0)$ scales with voxel sizes. Consequently, it is more informative to directly compute $S_{bb,off}(TE)-S_{bb,on}(TE)$. To analyze this, the voxel of spins is modeled as a magnetic dipole moment $\vec{m}$, and the resulting magnetic field $\vec{B_r}(\vec{r})$ at the receiver plane located a distance $r$ away is evaluated. Assuming the magnetization is uniform in the voxel of interest and the receiver coil is orthogonal to the transverse plane, the magnitude of $\vec{B_r}(\vec{r})$ simplifies to 
\begin{equation}
|\vec{B_r} (\vec{r})| = \frac{\mu _0}{4\pi}\frac{2|\vec{m}|}{r^3} = \frac{\mu _0}{4\pi}\frac{2M_o}{r^3}.
\label{eq13}
\end{equation}
\cite{griffiths2023introduction}. The induced flux on the receiver plane is now represented as 
\begin{equation}
\Phi = \int \vec{B_r} \cdot \hat{n} dA_s,
\label{eq14}
\end{equation}
where $A_s$ is the elementary sample area. Since $S_{bb}(0) \propto |\vec{B_r}|$, it follows that the signal scales linearly with voxel volume $V$. The blue curve shows the maximum detectable $\vec{B_r}$ field difference between current-on and current-off states, assuming the receiving plane is positioned 3 cm from the voxel of interest. The simulation results show that the maximum detectable field initially increases but eventually plateaus, indicating that increasing voxel size could potentially boost detectable contrast, but the effect doesn't scale linearly. In fact, continuing to increase the voxel size causes physiological noise to dominate rather than thermal noise, which is undesirable for the detection of contrast \cite{bodurka2007mapping}.

\subsubsection {Repetition Time (TR)}
The data rate of this encoding method is dictated by the TR. Because the data is binary-encoded, each implant transmits one bit per TR. However, the maximum available magnetization in each TR depends on the magnetization recovery from the previous TR. As illustrated in Fig. \ref{f6} (a), after the first RF excitation, the transverse magnetization $M_{xy}$ goes through $T_2^*$ decay, and the longitudinal component $M_z$ recovers with a time constant $T_1$. In subsequent TRs, the recovered $M_z$ serves as the initial transverse component after excitation. Consequently, shorter TR values result in less longitudinal recovery, leading to reduced SNR. This introduces a trade-off between data rate and CNR:  while shortening TR increases data rate, it also reduces contrast. In GRE, contrast can be maximized by choosing the Ernst angle as the flip angle to preserve transverse magnetization \cite{ernst1992nuclear}. However, this approach is not applicable in SE where the first RF pulse is typically 90$^\circ$. Fig. \ref{f6} (b) shows the decline in $C_n$ with decreasing TRs for current intensity 200 $\mu$A in a 10-turn 600 $\mu$m coil in SE-EPI when TE = 65 ms. The red line denotes the maximum achievable contrast under the same conditions when TR = $\infty$. 

Nevertheless, since the end goal is to maximize CNR, we can reduce noise by averaging the same data multiple times. For a data rate $f_{\text{data}}$, the period is $T_{\text{data}} = 1/f_{\text{data}}$. One bit of data can be acquired by setting TR = $T_{\text{data}}$, or repeating the same data multiple times at shorter TRs and averaging them. For instance, let N be the number of acquisitions repeated for the same data. If N = 2, TR = $T_{\text{data}}$/2, and similarly for higher values of N. This method has the potential to boost CNR by reducing noise by a factor of $\sqrt{N}$. Fig. \ref{f6} (c) plots the normalized CNR for N from 1 to 4 for data rate from 1/3 bps to 10 bps for the same setup as Fig. \ref{f6} (b). CNR of all scenarios is normalized to the SNR at time 0 of the received baseband signal. It can be observed that for data rate $<$ 0.6 bps, averaging helps boost CNR whereas at higher data rates, N = 1 yields the best CNR. In practical imaging scenarios involving multi-slice 2D acquisition of volumetric data, shortening TR does not reduce contrast because the longitudinal magnetization in each slice has more time to recover due to the interleaved nature of the acquisition process \cite{rieger2018time}.

\vspace{-1.25mm}
\subsubsection{Alignment}
\begin{figure}[t!]
\centering
\includegraphics[width=6cm]{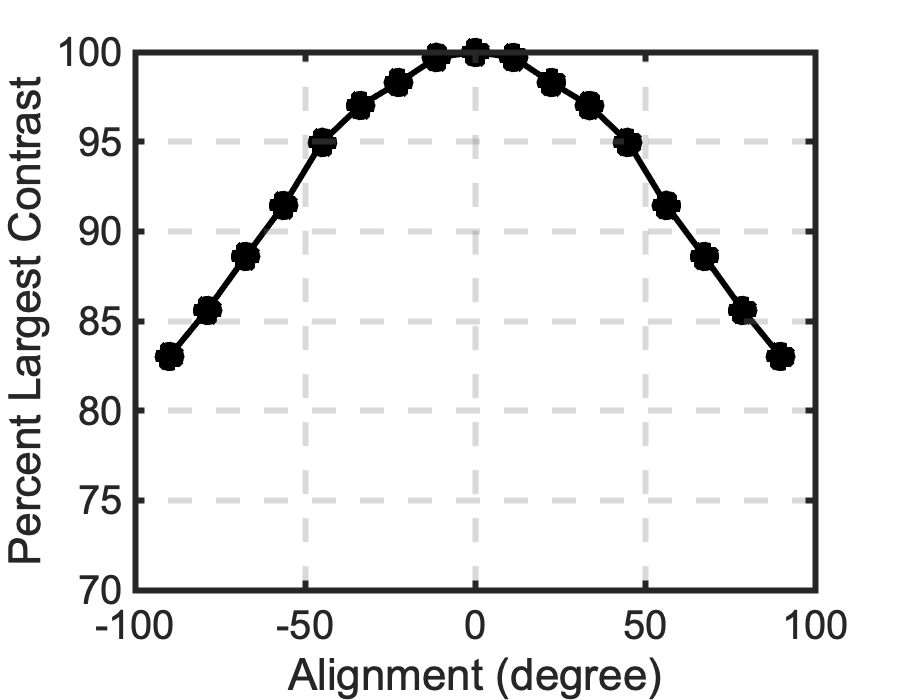}
\caption{The percentage of the normalized contrast to the normalized contrast at 0$^\circ$ when the micro-coil is aligned -90$^\circ$ to 90$^\circ$ away from Fig. \ref{f5} (a) \vspace{-2em}}
\label{f7}
\end{figure}

Fig. \ref{f7} demonstrates the method's robustness to device misalignment. All previous simulations assume the micro-coil is orthogonal to the $\vec{B_0}$ direction. However, it is essential to evaluate how the proposed method performs when the micro-coil is oriented at different angles. For a 10-turn 600 $\mu$m micro-coil with 100 $\mu$A current, we evaluate contrasts at rotation angles from -90$^\circ$ to 90$^\circ$ with 22.5$^\circ$ increments. The maximum achievable contrast occurs at 0$^\circ$. The maximum contrasts at all other rotation angles are normalized to the maximum contrast at 0$^\circ$. At the largest rotation angles (-90$^\circ$ and 90$^\circ$), 83\% of the maximum contrast can still be achieved. This approach is robust to misalignment because the modulation is created by field inhomogeneities caused by the induced magnetic field in the $\vec{B_0}$ direction. At all misalignment angles, the micro-coil is still able to generate significant inhomogeneity.

Based on the design considerations discussed above, the achievable CNR for a given micro-coil design in a particular MRI scanner can be estimated. For instance, the SNR of the SE-EPI sequence was experimentally measured on an MR750W GE3T MRI scanner (Waukesha, WI). SE-EPI scans were repeated 100 times on a coronal brain slice of a subject with voxel size 2$\times$2$\times$3.6 mm$^2$, TE = 35/65 ms, and TR = 1250 ms using a 17-channel commercial head coil. Noise images of the same field of view were acquired by turning off the RF excitation. The SNRs are respectively 121.06 and 72.94 for TE = 35 ms and 65 ms. A micro-coil design with 600 $\mu$m in width, 10 turns, 10 $\mu$m spacing, and 200 $\mu$A current was selected, along with a voxel size of 2$\times$2$\times$2 mm$^2$. Since the system is dominated by body noise, the noise floor remains the same regardless of voxel size. The measured SNR values were then scaled to estimate the effective noise at the chosen voxel dimensions. Based on these parameters, the estimated CNR at TE = 65 ms is 8.04.

\begin{figure}[t!]
\includegraphics[width=9cm]{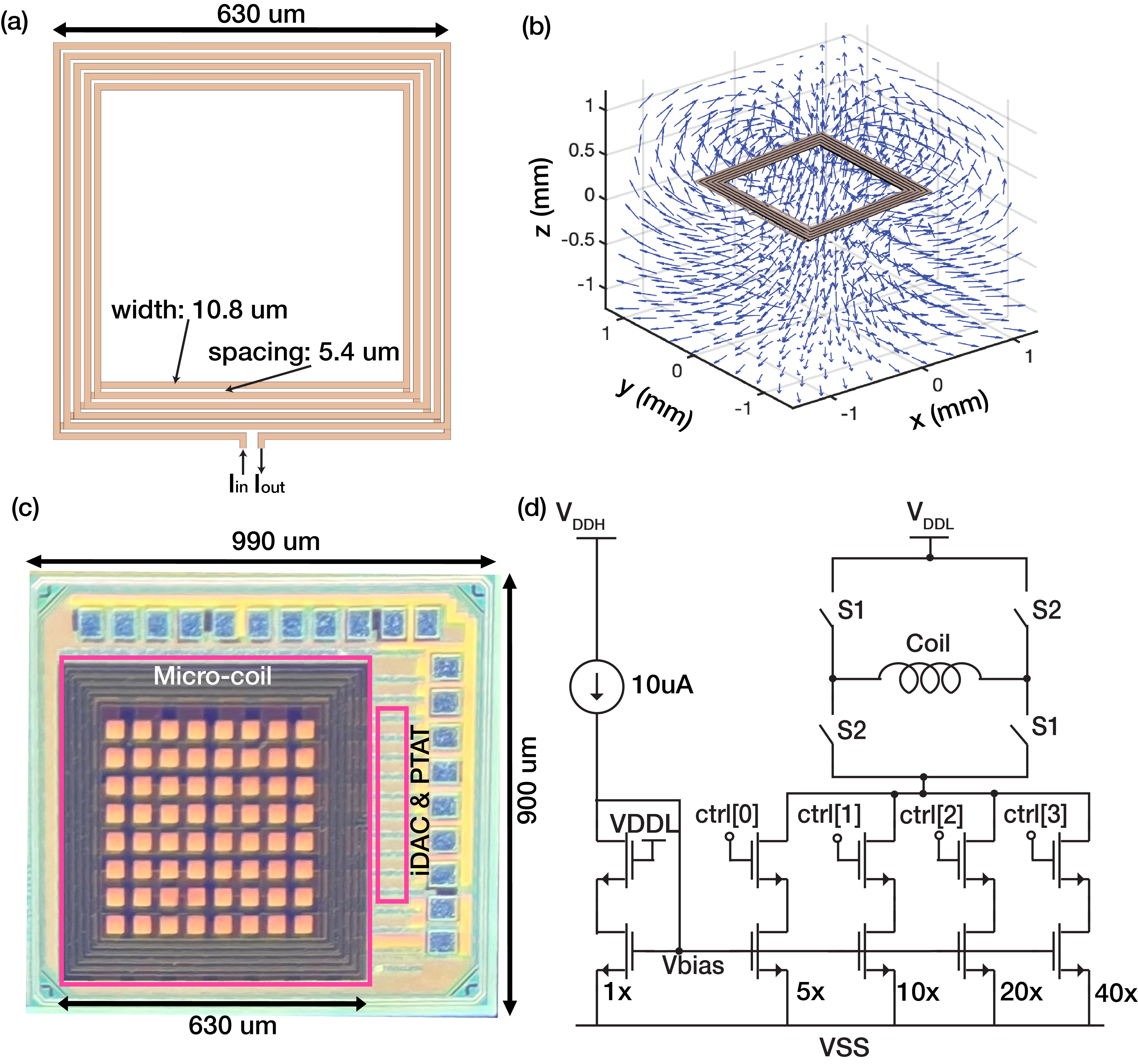}
\caption{(a) Top view of the 2-layer 10-turn micro-coil showing current flow directions, and (b) The magnetic field inhomogeneities it generates in a 2$\times$2$\times$2 mm$^2$ voxel. (c) Die micrograph of the chip. (d) Schematic diagram of the current DAC and H-bridge coil driver.\vspace{-2em}}
\label{chip_n_circuit}
\end{figure}

\section{Proof-of-concept Prototype Design}

\subsection{Integrated Circuit Implementation}
To validate the proposed method, a chip measuring 900$\times$990 $\mu $m$^2$ was designed and fabricated using the TSMC 28 nm process technology, as illustrated in Fig. \ref{chip_n_circuit} (c). The on-chip micro-coil occupies 630$\times$630 $\mu $m$^2$, with the remaining central area populated by dummies to satisfy design rule constraints. The micro-coil consists of  10 turns, with 5 turns on each of the top two thick metal layers connected by a via array. The design parameters of the micro-coil were chosen to achieve $C_n = 10\%$ at TE $= 35$ ms when driven by a 100 $\mu $A current. Under a $0.9 $V supply, this configuration results in a power consumption of 90 $\mu $W, which is achievable with several existing energy harvesting methods \cite{singer2021wireless}. The micro-coil layout was implemented in Cadence Virtuoso and subsequently ported into Ansys Maxwell 3D for electromagnetic simulation of the induced magnetic field inhomogeneity. The resulting field distributions within a 2$\times$2$\times$2 mm$^3$ voxel are plotted in Fig. \ref{chip_n_circuit} (b). 

Fig. \ref{chip_n_circuit} (d) shows the schematic diagram of the coil driver. Biphasic control of the coil current is achieved with an H-bridge, where S1 and S2 control the direction of the current based on the pulse sequence timing. To control the DC current level in the micro-coil, a proportional to absolute temperature (PTAT) circuit generates 10 $\mu$A of reference current under a 1.8V supply ($V_{DDH}$) supply. This reference is mirrored to a 4-bit current digital to analog converter (iDAC) ranging from 0 to 750 $\mu$A with 50 $\mu$A step size. The iDAC operates under a lower supply voltage of 0.9V ($V_{DDL}$).

\begin{figure}[t!]
\centering
\includegraphics[width=8.7cm]{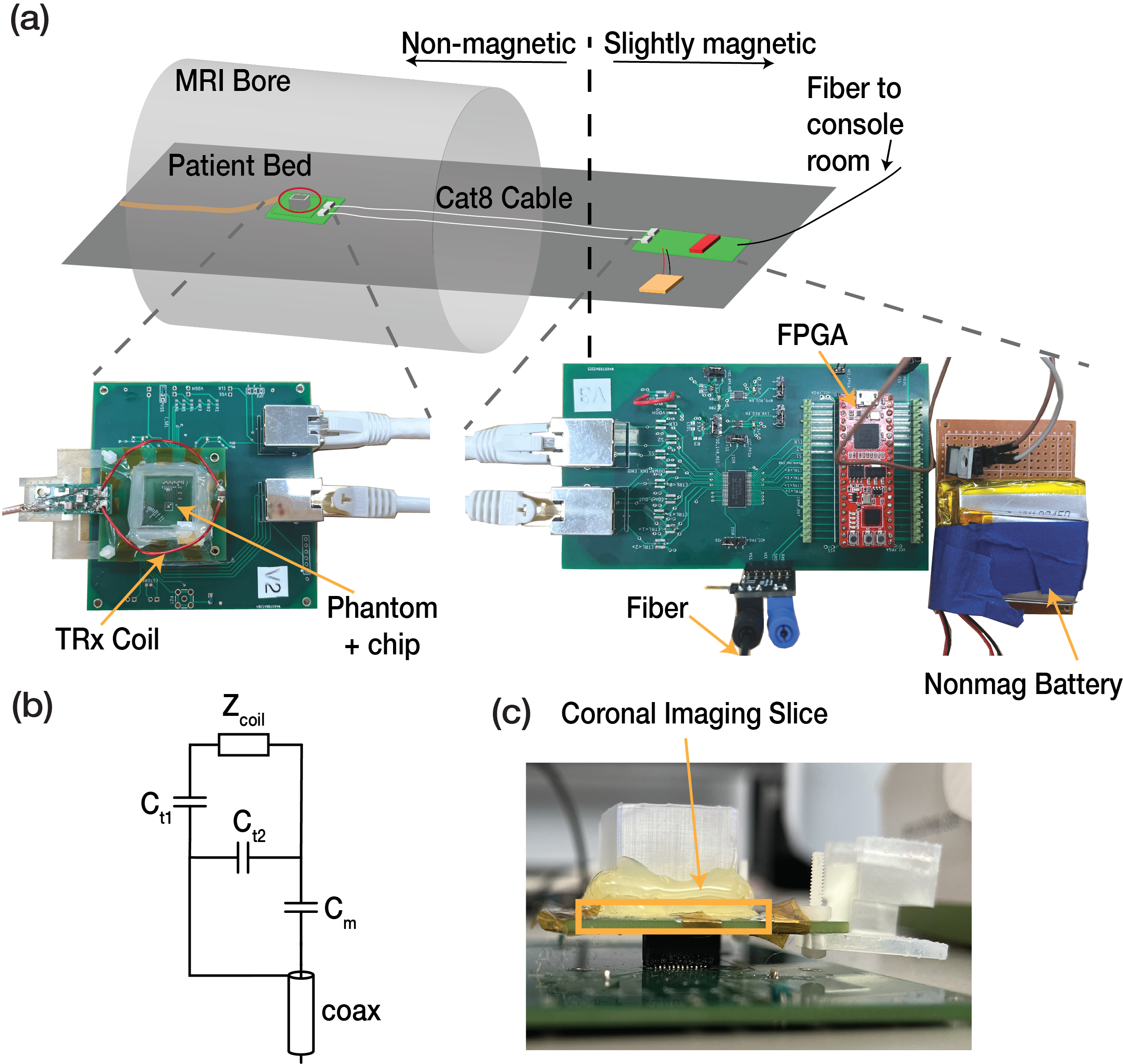}
\caption{ In-scanner in-vitro experiment setup: (a) Illustration of the implemented test setup. (b) Circuit diagram of the custom TRx coil. (c) Side view of the motherboard and daughterboard, with the coronal imaging slice highlighted.\vspace{-2em}}
\label{hardware_setup}
\end{figure}
\vspace{-2mm}
\subsection{Experiment Setup}
The chip was tested in the same GE3T scanner as in section III.B. The operation requires wired power delivery and digital control of the DAC current and switches, which were implemented using an external battery and a Fireant FPGA (Jungle Electronics). Testing wired electrical setups in the MRI environment involves several challenges: 1) The strong static magnetic field exerts translational forces on ferromagnetic materials, potentially pulling them into the magnet. 2) Ferromagnetic materials introduce substantial imaging artifacts and distortions. 3) RF magnetic field at the Larmor Frequency ($\sim$128 MHz for a 3T scanner), generated by 30 kilowatt-level amplifiers, can cause heating and induce currents in long metal traces and loops. The induced current could result in hardware malfunctioning. 4) Fast switching time-varying gradient fields (up to several kilohertz range) used in fast imaging sequences also induce current in hardware \cite{stafford2020physics}. 

To overcome these challenges, custom testing hardware was developed. First, to avoid the usage of ferromagnetic materials in the MR bore, the testing hardware was separated into two sections: nonmagnetic and slightly magnetic (but safe) (Fig. \ref{hardware_setup} (a)). In the nonmagnetic section, the chip was wire-bonded to a daughterboard with electro-palladium immersion gold (EPIG) surface finish to avoid the usage of nickel, which is a common yet magnetic material for PCB surface finishes. Clear epoxy was applied over the bond wires to protect them while exposing the micro-coil surface to generate maximum induced field inhomogeneity. The daughterboard was connected to a motherboard that contains the power supply and control signals. 

The magnetic section was located 2 meters away from the scanner bore to avoid magnetic force. It included a PCB containing a Fireant FPGA, voltage regulators,  logical level shifters, and a fiber optic breakout board (SparksFun) for receiving control signals from the scanner control room. The setup was powered by a nonmagnetic battery (PowerStream GM-NM103450). Power and control signals traveling between the magnetic and nonmagnetic sections were connected by two Cat8 cables, which are shielded twisted pairs to protect the long traces from RF and gradient effects. 

To synchronize the timing of current control, a transistor-transistor logic (TTL) signal indicating the presence of RF pulses was acquired from the scanner penetration cabinet. The electrical signal was converted to an optical signal using a fiber duplex modem (SparksFun) and sent via an optical fiber to the FPGA. Optical communication was chosen to avoid circuit malfunctioning caused by the gradient field. The FPGA processes the TTL signal, detecting the falling edges and controlling the current by synchronizing the S1 and S2 control signals on the chip with the scanner RF pulses. 

To mitigate RF-induced currents, a local phantom and transmit-receive (TRx) coil were fabricated. The chip was insulated using a layer of Saran wrap, over which a 3D-printed mold (2$\times$2$\times$1.5 cm$^3$) was affixed. The mold was filled with water doped with 3.3$\%$ $NiCl_2$ to mimic white matter tissue properties. A 4 cm diameter TRx coil made from magnet wire with a tuning and matching PCB was connected to the scanner via a preamp gateway box (Clinical MR Solutions LLC, Brookfield, WI) using a coaxial cable. 

It is important to note that these challenges are specific to this current prototype implementation. In a fully integrated MRDust system with wireless energy harvesting, the chip would be packaged as a miniature implant and placed in a larger tissue-mimicking phantom, eliminating the need for extended metal traces and substantially mitigating electromagnetic interference and imaging artifacts. 

\section{Results}

\begin{figure*}
\includegraphics[width=15cm]{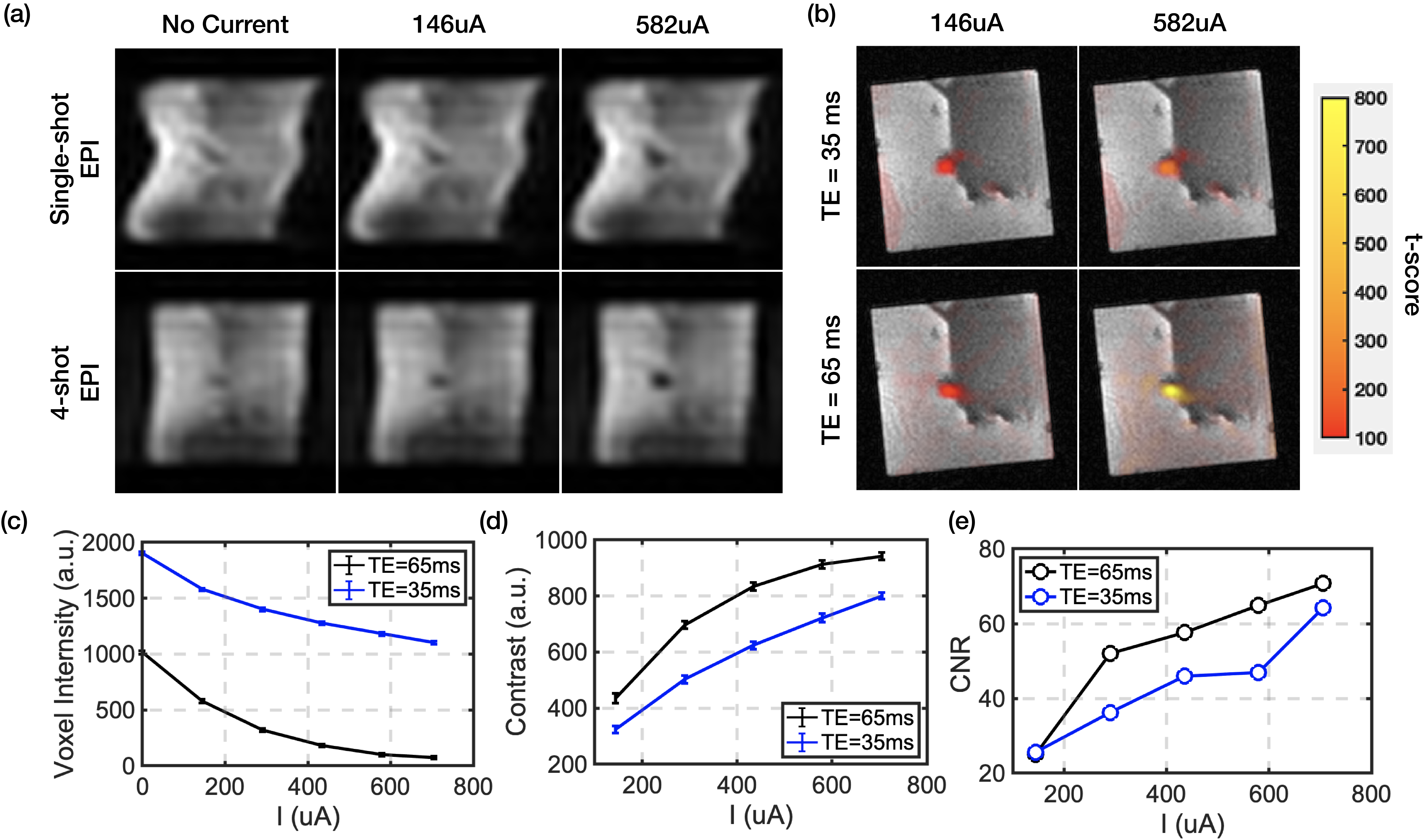}
\centering
\caption{(a) Single-shot and interleaved 4-shot SE-EPI coronal images acquired with 1.2$\times$1.2 mm$^2$ resolution, 3.6 mm slice thickness, 4$\times$4 cm$^2$ FOV, TR = 800 ms, TE = 65 ms. (b) Statistical analysis to identify regions of modulation: t-score values of each voxel are overlaid on a GRE scan with TE = 6.2 ms, TR = 150 ms, and flip angle 15$^\circ$. (c) The voxel with the largest signal change is selected based on t-test results and the average voxel intensity and standard deviation over different current levels and echo times are displayed. (d) For the same voxel as (c), the contrast between different current-on and current-off for different current levels is displayed for TE = 35 ms and 65 ms. (e) CNR for each current level in (d). \vspace{-2em}}
\label{fig9}
\end{figure*}

All results presented in this section were acquired on the coronal slice where the chip is located as shown in (Fig. \ref{hardware_setup} (c)). As demonstrated in \cite{zhao2023mrdust}, GRE-EPI sequences are highly susceptible to the $T_2^*$ effect, leading to significant signal loss. It is therefore difficult to acquire GRE-EPI images of the chosen slice at the desired TE since the imaging slice has significant field inhomogeneities caused by metal traces on the PCB. Alternatively, SE-EPI is more robust to such artifacts. Fig. \ref{fig9} (a) presents images acquired at various current levels using SE-EPI. The imaging parameters are:  1.2 $\times$ 1.2 mm$^2$ resolution, 3.6 mm thickness, 4 $\times$ 4 cm$^2$ FOV, TE = 65 ms, and TR = 800 ms. The top row displays single-shot EPI results, where each image is acquired in a single TR. It can be observed that a higher current value results in larger modulation. However, single-shot EPI images display geometric distortions, which are caused by the significant field inhomogeneities from the test setup \cite{schmitt2012echo}. A multi-shot SE-EPI sequence is less susceptible to such artifacts \cite{liao2019highly}, and 4-shot SE-EPI results of the same slice are displayed on the bottom row as a reference. With fully wireless hardware where PCB traces aren't present, the imaging results of single-shot EPI sequences would look similar to the bottom row. In these images, signal modulation remains localized to the region surrounding the chip.

To quantitatively study the location and intensity of signal modulation from the micro-coil, 50 images of the imaging slice were collected at various current levels for TE = 35 and 65 ms and TR = 1250 ms using single-shot EPI. The geometric distortions were corrected using the algorithm in \cite{jezzard1995correction}. A one-sided two-sample t-test was performed on each voxel with the hypothesis that a higher current level results in lower voxel intensity in the region of interest. The t-scores of the statistical analysis are overlaid on a GRE image of the same FOV acquired with TE = 6.2 ms, TR = 150 ms, and flip angle = 15$^\circ$ (Fig. \ref{fig9} (b)). The region of modulation can be clearly identified from the result. The voxel with the largest modulation was then identified by selecting the voxel with the lowest p-value. The intensities and standard deviations of the voxel at different TE and current values are plotted in Fig. \ref{fig9} (c). The corresponding contrast and CNR relative to when the current is off are shown in Fig. \ref{fig9} (d) and (e). It can be observed that longer TE results in larger modulation near the micro-coil local area. Even shorter TE leads to higher voxel intensities, longer TE provides overall greater contrast, and improves the CNR.
\vspace{-2mm}

\section{Conclusion and Discussion}
This work presents MRDust, an approach to integrate MRI with active sensor implants to realize a ``smart contrast agent" capable of encoding sensor data directly into MR images. The encoding is achieved by modulating MR voxel intensity with an actively driven on-chip coil. The application of a DC current through the coil induces local magnetic field perturbations, which in turn result in voxel amplitude modulation. To the best of the authors' knowledge, this work is the first to demonstrate active sensor data encoding via MRI for wireless sensor implants, while simultaneously enabling spatial localization and registration. The underlying physical principles were analytically described, and Bloch simulations were used to analyze the design trade-offs and establish guidelines for both micro-coil design and pulse sequence parameter selection. A prototype chip was fabricated and tested \textit{in vitro} in a clinical MRI system, demonstrating that a 10-turn 630 $\mu$m on-chip coil operating at 130 $\mu$W, when synchronized with an SE-EPI sequence, achieves a CNR of 25.58 at TE of 35 ms.

The MRDust approach offers a rather simple design method for data encoding and localizing for sensor implants. By using DC current, the system avoids the need for frequency synthesis and the tuning and matching of passive components typically required in conventional EM-based data encoding or localization, resulting in reduced area and simplified design. MRDust also integrates with existing MRI technology, eliminating the need for custom transceivers for communication and localization, an often significant overhead in clinical settings. Compared with the passive resonator in \cite{hai2019wireless}, this method has a much smaller form factor (\textless 1 mm) and more flexibility in sensor type and data transmission encoding. This method can also be easily extended to multiple implants in the same imaging slice. The spatial resolution of data readout is limited by the voxel size as two implants placed too close to each other may result in spatial crosstalk. This limitation can potentially be solved by increasing imaging resolution or using super-resolution algorithms \cite{shi2015lrtv}. 

\vspace{-1mm}
One limitation of the MRDust is its relatively low data rate. The approach is still well-suited for applications that don't require fast real-time data communication such as molecular sensing or physiological pressure sensing. For applications that do necessitate higher throughput, there are a number of ways to address the data rate limitation. For instance, recorded data can be temporarily stored on-chip and transmitted later at a slower rate, or alternatively, on-chip data compression or processing can be implemented to reduce the transmission burden.  In the future, this device may be integrated with wireless power delivery and sensor interfaces, and potentially scaled into miniaturized implant networks capable of multi-modal physiological data acquisition for both diagnostic and research applications.

\section*{Acknowledgment}

The authors thank the sponsors of Berkeley Wireless Research Center and the TSMC University Shuttle Program for chip fabrication, and professor Alp Sipahigil and Kadircan Godeneli for the technical discussions, and professor Brian Hargreaves from Stanford University for the Bloch Simulator code. This work was in part supported by the NIH under grants R01MH127104 and by the Apple PhD Fellowship in Integrated Systems.

\ifCLASSOPTIONcaptionsoff
  \newpage
\fi

\bibliography{refs}
\bibliographystyle{IEEEtran}

\end{document}